\documentclass[reprint,superscriptaddress,secnumarabic,amssymb, nobibnotes,noeprint, aps, prd,numerical]{revtex4-2}

\usepackage{graphicx}
\graphicspath{{images/}} 
\usepackage{siunitx}
\usepackage[colorlinks = true, linkcolor = black, urlcolor = blue, citecolor = black, anchorcolor = black]{hyperref}
\usepackage{xcolor}

\begin{document}

\title{Transparent Josephson Junctions in Higher-Order Topological Insulator WTe$_{2}$ via Pd Diffusion}%

\author{Martin Endres}
\email{martin.endres@unibas.ch}
\affiliation{Department of Physics, University of Basel, Klingelbergstrasse 82, 4056 Basel, Switzerland}

\author{Artem Kononov}
\affiliation{Department of Physics, University of Basel, Klingelbergstrasse 82, 4056 Basel, Switzerland}

\author{Michael Stiefel}
\affiliation{Laboratory for Nanoscale Material Science, Swiss Federal Laboratories for Material Science and Technology, EMPA, {\"U}berlandstrasse 129, 8600 D{\"u}bendorf, Switzerland}

\author{Marcus Wyss}
\affiliation{Swiss Nanoscience Institute, University of Basel, Klingelbergstrasse 82, 4056 Basel, Switzerland}

\author{Hasitha Suriya Arachchige}
\affiliation{Department of Physics and Astronomy, University of Tennessee, Knoxville, Tennessee 37996, USA}

\author{Jiaqiang Yan}
\affiliation{Department of Physics and Astronomy, University of Tennessee, Knoxville, Tennessee 37996, USA}
\affiliation{Material Science and Technology Division, Oak Ridge Laboratory,Oak Ridge, Tennessee 37831, USA}

\author{David Mandrus}
\affiliation{Department of Materials Science and Engineering, University of Tennessee, Knoxville, Tennessee 37996, USA}
\affiliation{Department of Physics and Astronomy, University of Tennessee, Knoxville, Tennessee 37996, USA}
\affiliation{Material Science and Technology Division, Oak Ridge Laboratory,Oak Ridge, Tennessee 37831, USA}

\author{Kenji Watanabe}
\affiliation{National Institute for Materials Science, 1-1 Namiki, Tsukuba 305-0044, Japan}

\author{Takashi Taniguchi}
\affiliation{National Institute for Materials Science, 1-1 Namiki, Tsukuba 305-0044, Japan}

\author{Christian Sch{\"o}nenberger}
\email{christian.schoenenberger@unibas.ch}
\affiliation{Department of Physics, University of Basel, Klingelbergstrasse 82, 4056 Basel, Switzerland}
\affiliation{Swiss Nanoscience Institute, University of Basel, Klingelbergstrasse 82, 4056 Basel, Switzerland}

\date{\today}%

\begin{abstract}
Highly transparent superconducting contacts to a topological insulator (TI) remain a persistent challenge on the route to engineer topological superconductivity.  Recently,  the higher-order TI WTe$_2$ was shown to turn superconducting when placed on palladium (Pd) bottom contacts, demonstrating a promising material system in pursuing this goal.  Here, we report the diffusion of Pd into WTe$_2$ and the formation of superconducting PdTe$_x$ as the origin of observed superconductivity. We find an atomically sharp interface in vertical direction to the van der Waals layers between the diffusion crystal and its host crystal, forming state-of-the-art superconducting contacts to a TI. The diffusion is discovered to be non-uniform along the width of the WTe$_2$ crystal, with a greater extend along the edges compared to the bulk. The potential of this contacting method is highlighted in transport measurements on Josephson junctions by employing external superconducting leads.  
\end{abstract}

\maketitle

\section{Introduction}
Topological insulators (TI) are insulating in the bulk while hosting gap-less boundary states in which the spin of the electron is locked to its momentum \cite{Fu2007}.  When brought in contact with a s-wave superconductor, a novel pairing mechanism is predicted with Cooper pairs that resemble an effectively spin-less superconductor \cite{Fu2008, Beenakker2013a}. Such topological superconductors could host Majorana bound states, the elementary building block of fault-tolerant quantum bits \cite{Hyart2013}. 

Fundamental to this approach is a highly transparent interface between the superconductor and topological insulator \cite{Schuffelgen2019} through which the boundary states are proximitized.  Even with state-of-the-art nano-fabrication it remains challenging to create such pristine material interfaces as oxidation \cite{Ye2016,Liu2015,Hou2020, Thomas2016}, contamination and rough crystal interfaces introduce defects and therefore decrease contact transparency \cite{Bagwell1992a}.  

The van der Waals (vdW) material WTe$_2$ is predicted to be a higher-order TI \cite{Schindler2018, Benalcazar2017, Wang2019a, Peng2017,Huang2020}, hosting topological edge states at its crystal hinges. It was recently shown that thin crystals of the material placed on top of palladium (Pd) bottom contacts turn superconducting \cite{Kononov2021a},  with a pronounced flow of super-current along the edges of a Josephson junction (JJ) formed out of this material system \cite{Kononov2020}.  

Here, we report diffusion of Pd into the WTe$_2$ forming superconducting PdTe$_x$ as the origin of superconductivity in the WTe$_2$/Pd system. The interface between PdTe$_x$ and WTe$_2$ in vertical direction to the vdW layers is found to be atomically sharp,  eliminating crystal-roughness between the superconductor and the higher-order TI completely. We further investigate the formation of PdTe$_x$ along the width of the WTe$_2$ host crystal and find it to be non-uniform, with greater extend along the edges compared to the bulk. The potential of this novel contacting method to WTe$_2$ is highlighted in transport measurements on JJs that show improved quality when contacted externally by an  intrinsic superconductor. 

\section{Pd Diffusion in WTe$_2$}{\label{Pd diff}}

\begin{figure}[]
    \centering
    \includegraphics[width=1\linewidth]{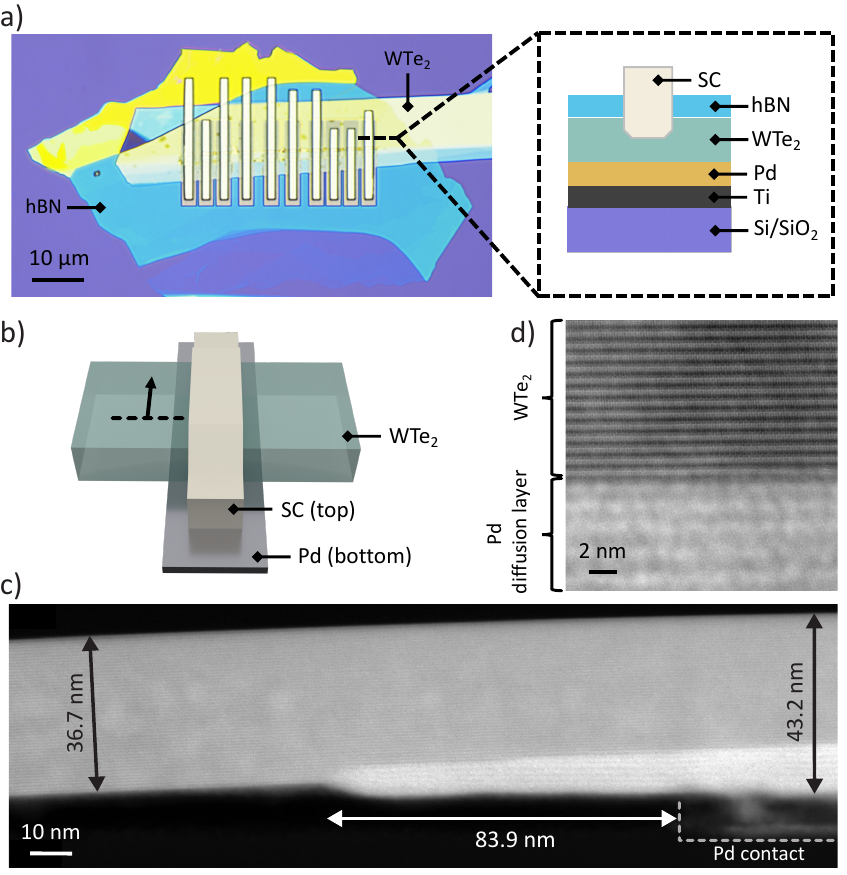}
    \caption{\textbf{Pd diffusion inside the WTe$_2$ crystal.} a) Left: optical image of the elongated WTe$_2$ flake, covered with hBN on top of Pd contacts, with additional superconducting contacts (niobium) on top. Right: a schematic cross section of the device in a region of single Pd contact. b) Illustration of the WTe$_2$ crystal on a single Pd bottom contact including a superconducting edge contact from the top. The direction of the cut lamella for the STEM image is indicated by the dashed line, the viewing direction of the image by the arrow. c) High resolution STEM image taken at the edge of the Pd bottom contact (indicated by grey dashed line at the bottom right). A bright diffusion layer at the interface between the Pd bottom contact and the WTe$_2$ crystal has formed. Black arrows indicate the thickness of the WTe$_2$ crystal, the white arrow shows the lateral extent of diffusion in WTe$_2$ from the edge of the original Pd contact. d) Zoom-in STEM image of the interface between the WTe$_2$ crystal and the diffusion layer.  }%
    \label{fig:1 Diffusion}
\end{figure} 

We begin with describing the general structure of WTe$_2$ Josephson junctions formed with Pd. Fabrication starts with patterning parallel lines of Pd on p-doped Si substrates with \SI{285}{\nano \meter} SiO$_2$ on top. Next, the vdW materials hexagonal boron nitride (hBN) and WTe$_2$ are exfoliated and afterwards stacked on top of the Pd bottom contacts, using a standard dry pick-up technique \cite{Zomer2014}. Until full encapsulation with hBN, WTe$_2$ is handled inside an inert glovebox-atmosphere to protect the material from oxidation \cite{Hou2020, Ye2016, Liu2015}.  After the stacking process, the polymer stamp is separated from the stacked device by heating the sample to \SI{155}{\degree C} for $\approx$~\SI{10}{min}.  The remaining polymer residues are chemically dissolved afterwards. Depending on the desired transport experiment, contact to WTe$_2$ is made either through the Pd bottom contacts or by etching through the covering top hBN and depositing superconducting contacts from the top.  It should be noted, that superconductivity is induced into WTe$_2$ by the Pd contacts alone \cite{Kononov2020, Kononov2021a} and that additional superconducting contacts are not required to form Josephson junctions. Fig.~\ref{fig:1 Diffusion}~a) demonstrates an optical image of one of the devices. This device was prepared specially for electron microscopy, so additional top superconducting contacts do not have any practical purpose and are placed to replicate real transport devices. An extended description of the fabrication process can be found in the supplementary materials \cite{supp, Blake2007, Zhao2015}. 

In order to investigate the origin of superconductivity in WTe$_2$ in contact with Pd, we conduct high resolution scanning transmission electron microscopy (STEM) imaging of the interface region. The illustration in Fig.~\ref{fig:1 Diffusion}~b) indicates the direction of the extracted lamella by a dashed line and the viewing direction by a perpendicular arrow.  Fig.~\ref{fig:1 Diffusion}~c) presents the STEM image taken with a high-annular angular dark field detector (HAADF) at the edge of a Pd bottom contact, reaching into the weak link of the junction.  Visible at first glance is a bright layer, that has formed at the interface between the Pd bottom contact and the WTe$_2$ crystal on top. Moreover, the original Pd contact in the bottom right corner of the Fig.~\ref{fig:1 Diffusion}~c) appears hollow and faded out, suggesting that the bright layer in WTe$_2$ is a result of Pd diffusion from the contact.

The presence of the diffusion layer in WTe$_2$ creates a pronounced swelling of the crystal, as highlighted by two thickness measurements of the WTe$_2$ flake in Fig.~\ref{fig:1 Diffusion}~c): inside the junction and on top of the Pd bottom contact. For pristine WTe$_2$ we extract an inter-layer spacing of $c\sim\SI{7.4}{\angstrom}$ that agrees with the literature value \cite{Brown1966, Chang2016}. Inside the diffusion layer the perceived layer spacing has doubled to $c\sim\SI{14.8}{\angstrom}$. We connect the change in the layer spacing with the formation of a new crystal structure at the interface of WTe$_2$ and Pd, rather than merely intercalation of the original crystal by Pd. The formation of a new structure is further supported  by fig.~\ref{fig:1 Diffusion}~d), where we see, that the transition between the newly formed crystal and WTe$_2$ is very sharp and takes place on a single layer scale. We also would like to note, that the diffusion forming the new structure is quite anisotropic. Laterally, along the vdW layers, the diffusion layer extends $\sim$ \SI{84}{\nano \meter} while vertically, perpendicular to the vdW layers, it only reaches $\sim$\SI{16}{\nano \meter} at its maximum. The lateral diffusion inside the JJ can diminish the length of the JJ, which could be especially prominent for the shorter junctions.

\begin{figure}[t!]
    \centering
    \includegraphics[width=1\linewidth]{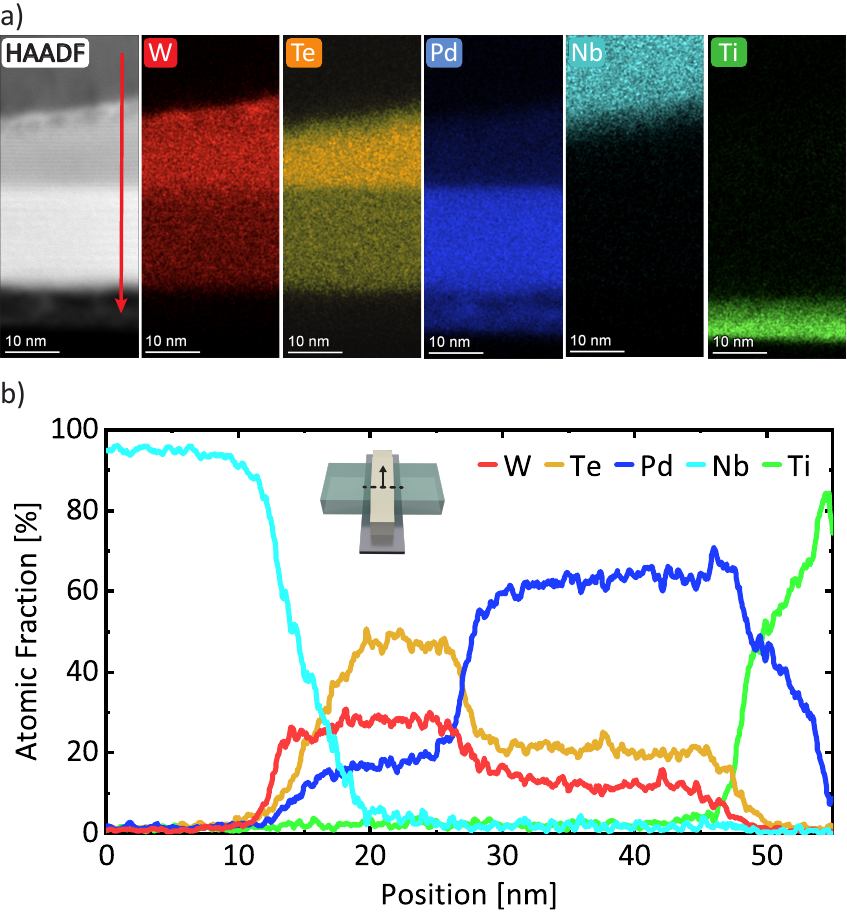}
    \caption{\textbf{EDX analysis of the Pd diffusion.} a) STEM image with the direction of the line cut in b) indicated by a red arrow. Presented towards the right is the EDX analysis with elements existent in the device.  Moving from the bottom to the top, the Pd bottom contact is followed by highly Pd interspersed WTe$_2$ layer that has structurally changed.  Above, the crystal transitions sharply into the original crystal structure.  b) EDX line cut along the direction indicated in a), with the position of the investigated lamella marked in the insert. Pd has diffused in vertical direction through the entire WTe$_2$ crystal,  with a sharp concentration increase in the structurally changed area that coincides with the STEM image.}%
    \label{fig:2 EDX}
\end{figure} 

In the next section we analyze the atomic composition of the diffusion layer using energy dispersive x-ray (EDX) analysis. Fig.~\ref{fig:2 EDX}~a) on the left shows a STEM image taken at the position of a superconducting niobium (Nb) top-contact in this device. For better orientation, the location is illustrated in b).  From the bottom to the top, the faded Pd bottom contact, the Pd diffusion layer adjacent to the pristine WTe$_2$ crystal and the Nb top contact are visible. Towards the right, EDX spectra of the elements in this slab are shown. Nb (turquoise) and the sticking layer for the Pd bottom contacts, titanium (Ti) are at their expected positions. Qualitatively, the concentration of tungsten (W) and tellurium (Te), represented in red and orange, respectively, are maximal in the unchanged WTe$_2$ crystal but reduced in the diffusion layer. Pd (in blue) has diffused through the entire WTe$_2$ crystal and is the dominating element inside the structurally changed layer. The concentration of Pd at the position of the original bottom contact is diminished, suggesting that depletion of the available material stopped the further growth of the diffusion layer. 

A quantitative analysis of the crystal composition is shown in Fig.~\ref{fig:2 EDX}~b), following a trace indicated by the red arrow in the STEM image in a). The ratio of W:Te is $\sim$ 1:2 and remains the same throughout most of the thickness.  For Pd, two distinct concentration levels are visible, a high level $\sim\,60\%$ that coincides with the structurally changed lattice and a second, low level $\sim\,20\%$ inside the preserved WTe$_2$ crystal. The ratio of Pd:Te $\sim$ 3:1 suggests that the diffusion layer is not one of the known superconductors PdTe or PdTe$_2$ \cite{Karki2012, Tiwari2015,  Das2018, Voerman2019}. In the unchanged WTe$_2$ crystal above, Pd likely intercalates the vdW layers. So possibly, a threshold concentration of Pd is required to trigger the crystallographic change, such that the vertical extend of the diffusion layer is determined by the interplay of available Pd and thermal activation energy. The remaining Pd concentration above the PdTe$_x$ layer quickly decays in direction parallel to the vdW layers and extends $\sim$ \SI{50}{\nano \meter} laterally beyond the structurally changed diffusion layer, as shown in \cite{supp}.

\section{Diffusion along the Edges}

\begin{figure}[]
    \centering
    \includegraphics[width=1\linewidth]{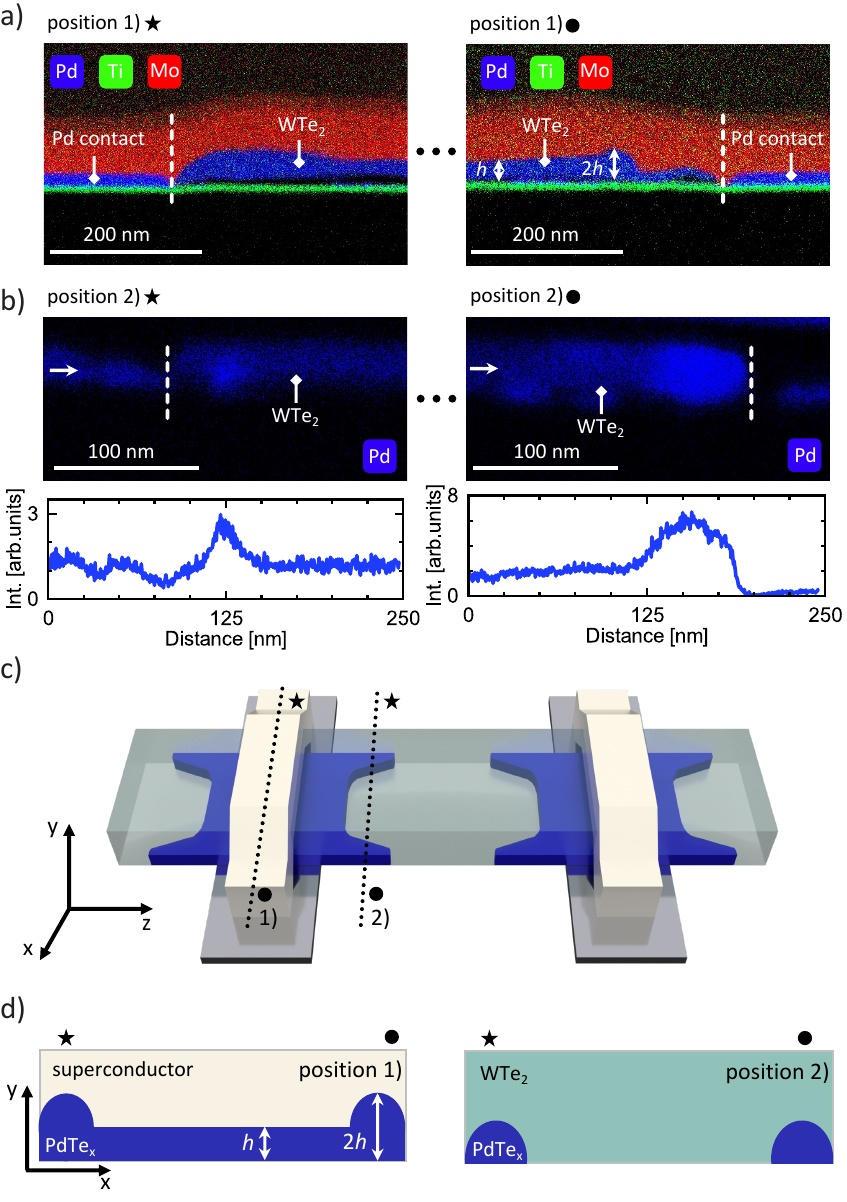}
   \caption{\textbf{Enhanced Pd diffusion along the edges of the WTe$_2$ crystal.} 
a) EDX analysis of a cross section taken on the Pd contact along position 1), as indicated in the schematics in c). The given sample was equipped with a superconducting MoRe contact, evidenced by the Mo EDX signal in red. The left and right image correspond to the crystal edges marked by $\star$ and $\bullet$ in the schematics, respectively.  Visible is the swelling of WTe$_2$ to a thickness $\sim 2h$ at the edge, compared to the bulk thickness $\sim h$, indicated in the right image. b) EDX signal and extracted intensity (Int.) profile taken towards the inside of the junction, indicated by position 2) in c).  The left and right image correspond to positions $\star$ and $\bullet$ of the WTe$_2$ crystal, respectively.  Visible in the EDX data is an increased intensity of the Pd signal at the edges of the crystal compared to the bulk.  The increased Pd concentration is as well visible by the enhanced EDX signal in the line cuts taken along the direction pointed out by the horizontal arrow. 
c) Illustration of the inhomogeneous diffusion profile of PdTe$_{x}$ inside the WTe$_{2}$ host crystal.  The self-formed PdTe$_x$ layer is drawn in blue inside the host crystal. 
d) Cross sectional cuts through the illustration along positions 1) and 2) in c). The edges are marked by $\star$ and $\bullet$ for orientation.}%
    \label{fig:3 Edge Diffusion}
\end{figure} 

Further, we investigate the uniformity of the PdTe$_x$ diffusion layer along the width of the Josephson junction. For this, we have analysed several lamellas that are oriented perpendicular to the direction of current in the JJ. The regions near the physical edges of WTe$_2$ are of particular interest, since additional Pd is available there due to the Pd bottom contacts extending beyond the crystal.

The first lamella was cut out through the middle of the bottom Pd contact in a sample with additional, this time molybdenum-rhenium~(MoRe), top contacts, as illustrated by position 1) in Fig.~\ref{fig:3 Edge Diffusion}~c).  Fig.~\ref{fig:3 Edge Diffusion}~a) presents two EDX spectra obtained at the two edges marked by $\star$ and $\bullet$ in the previous illustration. Outside of the WTe$_2$ flake we observe a layer of Pd with uniform thickness sandwiched between the MoRe top- and Ti bottom-layer. Interestingly, inside WTe$_2$ near the edges, the thickness \textit{h} of the PdTe$_x$ diffusion layer increases within a region of $\sim100$~nm away from the edge, as marked in the right spectrum of Fig.~\ref{fig:3 Edge Diffusion}~a). Further from the edges,  Pd is evenly distributed throughout the WTe$_2$ crystal. The difference of PdTe$_x$ thickness on the edges and in the middle of WTe$_2$ reaches a factor of $\sim$ 2. 

The increase in thickness of PdTe$_x$ near the edges can be intuitively explained taking into account the fabrication procedure. During the last step of stacking, when the substrate is heated up to \SI{155}{\degree C} to release the hBN/WTe$_2$ stack, the formation of the PdTe$_x$ takes place. In WTe$_2$ far away from the edges this process stops before reaching the full thickness of the flake due to depletion of available Pd. Near the edges, due to availability of additional Pd, this process continues potentially even through the whole thickness of WTe$_2$. Afterwards, the top WTe$_2$ layers, not transformed to PdTe$_x$, are etched away during CHF$_{3}$/O$_{2}$ plasma etching of hBN prior to the deposition of the superconductor. This explanation is further corroborated by the uniform Pd concentration in Fig.~\ref{fig:3 Edge Diffusion}~a) in contrast with the step in concentration in Fig.~\ref{fig:2 EDX}~b).

The increased Pd availability on the edges of the WTe$_2$ has the potential not only to increase the PdTe$_x$ thickness, but also to provide further diffusion inside the Josephson junction. To check this, we made a second lamella from another sample, which is cut inside the Josephson junction close to the end of Pd bottom contact, as shown as position 2) in Fig.~\ref{fig:3 Edge Diffusion}~c). Visible in the EDX spectrum in Fig.~\ref{fig:3 Edge Diffusion}~b) is an elevated intensity of Pd compared to the bulk at both ends of the crystal, highlighted by line cuts through the spectra along the horizontal arrows. This indicates, that PdTe$_x$ is indeed penetrating further inside the junction along the edges of WTe$_2$ as visualized in Fig.~\ref{fig:3 Edge Diffusion}~c).  

We can roughly estimate the extent of the PdTe$_x$ diffusion along the edges. Assuming that an increase by a factor of 2 in thickness \textit{h} of PdTe$_x$ on the edges, as compared to the bulk (see Fig.~\ref{fig:3 Edge Diffusion}~d)), yields the same increase in the diffusion inside the JJ along the edge. Taking from fig.~\ref{fig:1 Diffusion}~c), that the PdTe$_x$ layer extends $\sim 85$~nm inside the junction in the bulk, we would expect it to extend  $\sim 170$~nm along the edges. This diffusion could generate signatures of `artificial' edge supercurrents not due to a topological state. Nonetheless, evidence of topological hinge states in WTe$_2$ has been observed in combination with superconducting niobium contacts \cite{Choi2020, Huang2020}, where no edge diffusion is expected.

\section{Josephson junction with fully superconducting contacts}

\begin{figure*}[htp]
    \centering
    \includegraphics[width=1\linewidth]{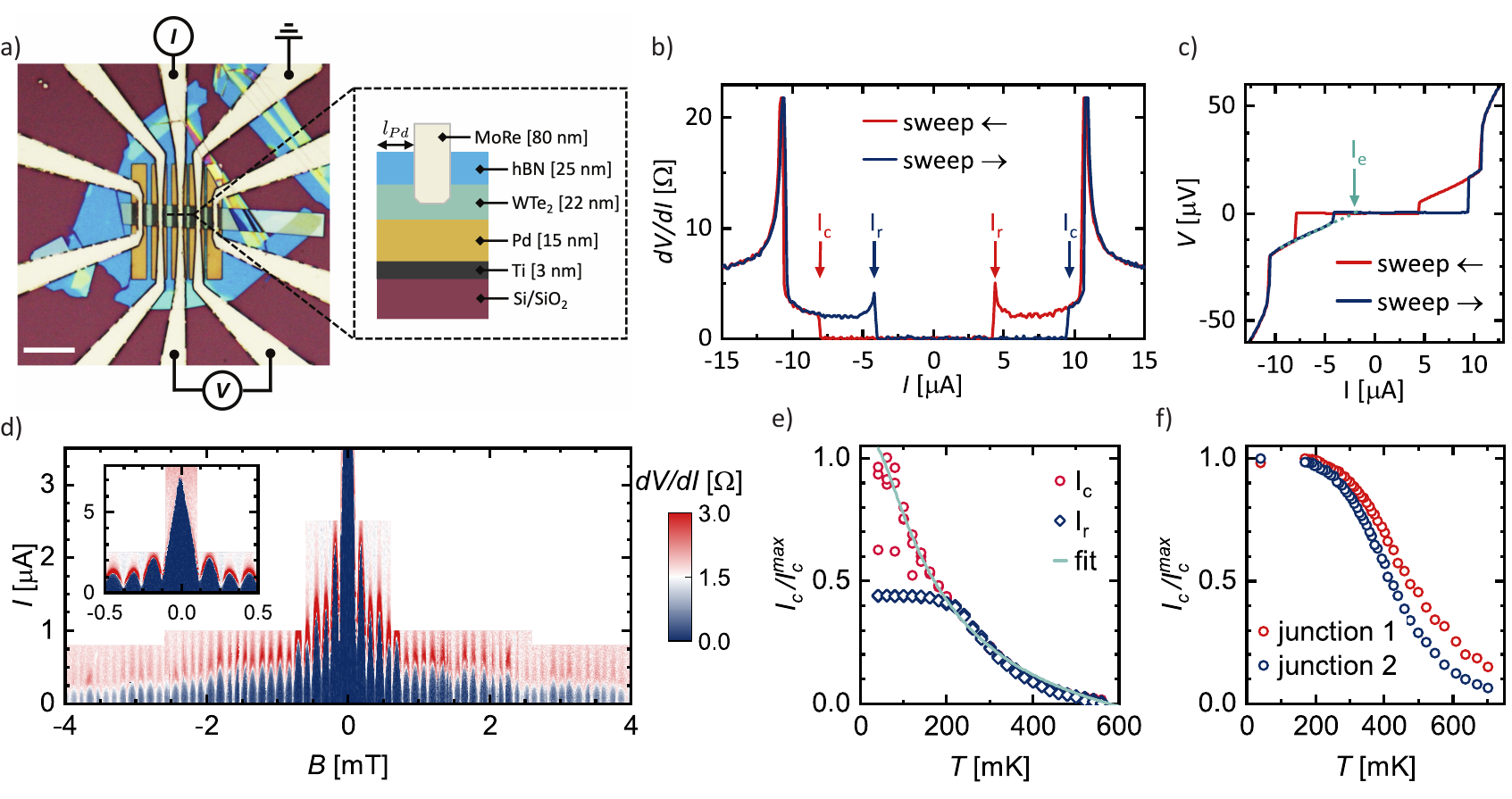}
    \caption{\textbf{Switching characteristics of Josephson junctions with superconducting or normal leads.} a) Optical image of the device with an illustration of the quasi four-terminal measurement setup.  The fabricated layer sequence is shown on the right. Scale bar is \SI{10}{\micro \meter}. b) $dV/dI$ curves at zero magnetic field of a JJ with superconducting contacts for two different sweep directions of the bias current. The hysteretic switching current depending on the sweep direction is visible from the shift of the critical current $I_{c}$ and the re-trapping current $I_{r}$.  c) $V(I)$ curves corresponding to the data in b). The excess current $I_{e}$ is evaluated from the intersection of the extended $V(I)$ curve to zero voltage. d) $dV/dI$ as a function of bias current $I$ and perpendicular magnetic field $B$, following the `Fraunhofer' pattern expected for a Josephson junction. The insert shows the full range of the central lobe around $B=0$. e) Temperature dependence of the switching and re-trapping currents $I_{c}$ and $I_{r}$, respectively, of the same device as in b).  The data $I_{c}(T)$ is fitted in the scope of a diffusive long junction, plotted in light green. f) Switching current $I_{c}$ as a function of temperature $T$ for two different JJs that are contacted only through Pd bottom contacts. }%
    \label{fig:Temperature Switching}
\end{figure*} 

During the formation of PdTe$_x$ in WTe$_2$ the majority of the Pd from the bottom contacts is depleted (see Fig.~\ref{fig:1 Diffusion}-\ref{fig:3 Edge Diffusion}), thus creating a low quality interface between bottom contacts and the newly formed Josephson junction. In this section we demonstrate a method to harness the full potential of the high quality Josephson junction formed in WTe$_2$ with Pd diffusion by employing additional superconducting contacts from the top.

The fabrication process follows the description in section~\ref{Pd diff}. After obtaining the stack, superconducting leads are patterned via standard e-beam lithography and sputtered onto the sample after etching through the top hBN with CHF$_{3}$/O$_{2}$ plasma. Prior to the deposition of MoRe superconducting leads, we perform a short Ar milling inside the sputtering chamber to remove the oxide layer from WTe$_2$.  In order to avoid degradation of the JJs due to etching, the superconducting top contacts are separated by a distance $l_{Pd} \sim$ \SI{0.5}{\micro \meter} from the edge of the Pd bottom contacts, as indicated in the schematics in Fig.~\ref{fig:Temperature Switching}~a). Fig.~\ref{fig:Temperature Switching}~a) shows a finished device with top MoRe leads and its fabricated layer sequence to the right.

Measurement of the device is performed in a quasi four-terminal setup,  illustrated in Fig.~\ref{fig:Temperature Switching}~a). In this configuration, the measured differential resistance includes the contribution from the Josephson junction and the resistances of the interfaces between the MoRe and superconducting PdTe$_x$, but excludes the line resistances in the cryostat.  Fig.~\ref{fig:Temperature Switching}~b) and c) show the $dV/dI(I)$ and $V(I)$ dependencies, measured on a \SI{1}{\micro\meter} long Josephson junction, with their behavior being representative for a number of samples we have studied. The curves reveal several abrupt transitions with current. Steps at $I\sim\pm$~\SI{11}{\micro\ampere} have minimal hysteresis and correspond to the switching of superconducting PdTe$_x$ to the normal state or alternatively to a Josephson junction that has potentially formed at the interface of the vdW stack with MoRe \cite{Sinko2021}. 

The observed vanishing resistance at low bias currents by itself is not sufficient to ensure that the fabricated device performs indeed as a JJ. Potentially, the inhomogeneous diffusion of PdTe$_x$ could lead to a closed superconducting path through the weak link.  In order to rule out this option, we study the dependence of $dV/dI$ on the bias current $I$ and perpendicular magnetic field $B$, as shown in Figure~\ref{fig:Temperature Switching}~d). Visible is a periodic `Fraunhofer'-like interference pattern that is a key signature of the Josephson effect. The oscillation periodicity $\delta B =$ \SI{0.13}{\milli \tesla} is connected to a flux quantum $\Phi_{0}$ threading the effective junction area $A_{eff} = w \times \ell_{eff}$,  with $w =$ \SI{4.3}{\micro \meter} being the width of the junction and $\ell_{eff}$ being the effective length. The calculated $\ell_{eff} \sim$ \SI{3.8}{\micro \meter} exceeds the physical junction length of $\sim$\SI{1}{\micro \meter}. However, it can be explained by the contact geometry, assuming that half of the magnetic flux through the superconducting contacts is screened into the junction \cite{Ghatak2018anomalous}.  Additionally, a close look at the amplitude of consecutive lobes reveals a non-monotonous behavior, reminiscent of an even-odd effect.  A non-sinusoidal current phase relation of the junction \cite{Kononov2020} or an inhomogeneous current distribution \cite{Ghatak2018anomalous},  originating from the diffusion profile of PdTe$_x$, can create this feature.  

Having established the Josephson effect through WTe$_2$, we take a closer look at the lower current behavior observed in fig.~\ref{fig:Temperature Switching}~b) and c). First, in the superconducting branch of the JJ, the differential resistance is zero, implying that there is no measurable contribution of the MoRe/PdTe$_x$ interfaces. Second, in contrast to the previously studied devices with solely Pd leads~\cite{Kononov2020}, the switching behavior is highly hysteretic. The transition from the superconducting to the resistive branch, denoted by the switching current $I_c$, takes place at higher absolute current values than the transition in opposite sweep direction, denoted by the re-trapping current $I_r$, highlighted in Fig.~\ref{fig:Temperature Switching}~b).

Even though the hysteretic switching of the Josephson junction is most commonly explained by the junction being in the underdamped regime~\cite{Tinkham1994}, we would argue that in our case overheating~\cite{DeCecco2016, Courtois2008} plays the dominating role. Starting from the superconducting branch, no heat is dissipated in the Josephson junction before switching to the resistive branch. In contrast, lowering the bias current from the resistive branch includes dissipation of heat in the normal weak link, leading to a higher electron temperature. This explanation is corroborated by the temperature dependence of $I_{c}$ and $I_{r}$ shown in Fig.~\ref{fig:Temperature Switching}~e). Moving from high towards low temperatures, $I_{c}$ and $I_{r}$ both increase continuously down to $T \sim$ \SI{220}{\milli \kelvin}, when $I_{r}$ begins to saturate while $I_{c}$ remains increasing. Furthermore, this explanation is additionally supported by the devices with only Pd contacts. There, due to the normal Pd contacts remaining dissipative at all times, $I_{c}$ saturates at low temperatures, as shown in Fig.~\ref{fig:Temperature Switching}~f). 

Next, we characterize the quality of the PdTe$_x$/WTe$_2$ interface.  Compared to conventional superconducting contacts to WTe$_2$, the Josephson effect in junctions formed by Pd interdiffusion is found to be more robust in terms of junction length and magnetic field resilience \cite{Kononov2020}. Additionally, due to reduced heating effects, the here proposed devices support a critical current density twice as large compared to conventional contacts \cite{Choi2020} at four times the junction length.  The \SI{1}{\micro \meter} long junction presented in fig.~\ref{fig:Temperature Switching}~b) maintains the Josephson effect with a critical current density $j_{c}$ of up to $j_{c} >$ \SI{e8}{\ampere \per \meter \squared} while comparable junctions with even a shorter length up to \SI{230}{\nano \meter} and conventional superconducting contacts are found to be limited by $j_{c} \sim$ \SI{1e7}{\ampere \per \meter \squared} -  \SI{5e7}{\ampere \per \meter \squared} \cite{Choi2020}.

Further,  from fig.~\ref{fig:Temperature Switching}~e) we see that $I_{c}$ is suppressed at \SI{0.6}{\kelvin}, which is lower than the critical temperature $T_{c} =$ \SI{1.2}{\kelvin} \cite{Kononov2021} of the formed PdTe$_{x}$. We connect this reduction with the great length of the junction and fit for this reason $I_{c} (T)$ with an expression for a long diffusive junction \cite{Dubos2001, DeCecco2016}

\begin{equation}
I_{c}=\eta \frac{a E_{Th}}{e R_{N}} \left[1-b \; exp\left(\frac{-a E_{th}}{3.2 k_{B}T}\right)\right].
\end{equation}

Here, $a$ and $b$ are constants equal to 10.82 and 1.30, respectively,  $k_{B}$ the Boltzmann constant and $E_{Th}$ the Thouless energy. The empirical pre-factor $\eta \in [0,1]$ can be interpreted as a measure for the interface quality, scaling the maximum $I_{c}$. The data in fig.~\ref{fig:Temperature Switching}~e) is well described by the model which yields $\eta =$ 0.5 and $E_{Th} =$ \SI{3.87}{\micro e \volt}. A similar fit procedure for the data in fig.~\ref{fig:Temperature Switching}~f), obtained from a junction with only Pd contacts, is not reliable as $I_{c}(T<400\;\text{mK})$ is limited by heating effects and deviates strongly from the theoretical prediction.

The evaluation of the interface transparency is corroborated by the excess current $I_{e}$, extracted from the $V(I)$ curve in fig.~\ref{fig:Temperature Switching}~c). We extrapolate $I_{e}$ after the transition from the superconducting to the resistive branch and obtain \cite{Bai2020} $I_{e}R_{N}/\Delta \sim$ 0.03, using $\Delta = 1.76 k_{B} T_{c} = $ \SI{182}{\micro e \volt} \cite{Kononov2021}. In the framework of Octavio-Tinkham-Blonder-Klapwijk theory \cite{Octavio1983, Flensberg1988subharmonic}, this relates to a junction transparency $T = 1/(1+Z^{2})\sim$ 0.5, with $Z \sim$ 1.1.

At this point we would like to comment on the role of $R_{N}$ in the two analytical models of the preceding analysis yielding a transparency $\sim$ 0.5. The bulk conductivity of WTe$_2$ increases with flake thickness \cite{Xiang2018}. These additional bulk modes in the normal state shunt $R_{N}$, but do not participate in superconducting transport of the long junction due to their fast decay $I_{c,bulk} \propto \ell_{mfp}^2/L_{w}^{3}$, compared to ballistic edge modes $I_{c,edge} \propto 1/L_{w}$ \cite{Murani2017}, with $\ell_{mfp}$ being the electronic mean free path. This phenomenon has also been reported in JJs formed of the topological material Bi$_2$Se$_3$ \cite{Galletti2014}. The extracted transparency is therefore systematically underestimated and serves only as a lower bound to the real value.

\section{Conclusion}
We have demonstrated a robust method to form atomically sharp superconducting contacts to WTe$_2$ mediated by Pd diffusion during stacking.  Josephson junctions formed by these contacts are highly transparent. Given recent reports of similar processes in BiSbTe~\cite{Bai2020, Bai2022} this method could be a promising approach for other topological candidates based on Te compounds. We have further demonstrated that the diffusion inside the host crystal could be non-uniform, generating false signatures of superconducting edge currents. Therefore, caution has to be exercised in the evaluation of diffusion driven Josephson junctions, when assigning it to topological superconductor. Furthermore, we proposed a method to avoid overheating in transport through Pd diffusion mediated Josephson junctions by employing additionally superconducting leads.

\textbf{Note.} During the preparation of this manuscript we became aware of a recent publication~\cite{Ohtomo2022}, which also demonstrates the interdiffusion of Pd into WTe$_2$ with the formation of PdTe leading to superconductivity.
\\

\section*{Acknowledgments}
We thank Paritosh Karnatak for fruitful discussions.
This project has received funding from the European Research Council (ERC) under the European Union’s Horizon 2020 research and innovation programme: grant agreement No 787414 TopSupra, by the Swiss National Science Foundation through the National Centre of Competence in Research Quantum Science and Technology (QSIT), and by the Swiss Nanoscience Institute (SNI).
A.K. was supported by the Georg H.~Endress foundation.
D.G.M. and J.Y. acknowledge support from the U.S. Department of Energy (U.S.-DOE), Office of Science - Basic Energy Sciences (BES), Materials Sciences and Engineering Division.
H.S.A. was supported by the Gordon and Betty Moore Foundation's EPiQS Initiative through Grant GBMF9096and the Shull Wollan Center Graduate Research Fellowship.
D.G.M. acknowledges support from the Gordon and Betty Moore Foundation’s EPiQS Initiative, Grant GBMF9069.
K.W. and T.T. acknowledge support from the Elemental Strategy Initiative conducted by MEXT, Japan and the CREST (JPMJCR15F3), JST.

\section*{Author Contributions}
M.E. has fabricated the devices. M.E. and A.K. measured the devices in transport. M.S. and M.W. performed the STEM and EDX imaging. H.S.A., J.Y. and D.G.M. provided the WTe$_2$ crystals. K.W., T.T. provided hBN crystals. M.E., A.K. and C.S.  analysed the data and wrote the manuscript. 

\section*{Competing Interests}

The authors declare no competing interest. 

\section*{Data availability}
All data in this publication are available in numerical form in the Zenodo repository at \url{https://doi.org/10.5281/zenodo.6556998}.

\bibliography{bib_main}

\end{document}


\title{Supplementary Information: Transparent Josephson Junctions in Higher-Order Topological Insulator WTe$_{2}$ via Pd Diffusion
}%

\author{Martin Endres}
\affiliation{Department of Physics, University of Basel, Klingelbergstrasse 82, 4056 Basel, Switzerland}

\author{Artem Kononov}
\affiliation{Department of Physics, University of Basel, Klingelbergstrasse 82, 4056 Basel, Switzerland}

\author{Michael Stiefel}
\affiliation{Laboratory for Nanoscale Material Science, Swiss Federal Laboratories for Material Science and Technology, EMPA, {\"U}berlandstr. 129, 8600 D{\"u}bendorf, Switzerland}

\author{Marcus Wyss}
\affiliation{Swiss Nanoscience Institute, University of Basel, Klingelbergstrasse 82, 4056 Basel, Switzerland}

\author{Hasitha Suriya Arachchige}
\affiliation{Department of Physics and Astronomy, University of Tennessee, Knoxville, Tennessee 37996, USA}

\author{Jiaqiang Yan}
\affiliation{Department of Physics and Astronomy, University of Tennessee, Knoxville, Tennessee 37996, USA}
\affiliation{Material Science and Technology Division, Oak Ridge Laboratory,Oak Ridge, Tennessee 37831, USA}

\author{David Mandrus}
\affiliation{Department of Materials Science and Engineering, University of Tennessee, Knoxville, Tennessee 37996, USA}
\affiliation{Department of Physics and Astronomy, University of Tennessee, Knoxville, Tennessee 37996, USA}
\affiliation{Material Science and Technology Division, Oak Ridge Laboratory,Oak Ridge, Tennessee 37831, USA}

\author{Kenji Watanabe}
\affiliation{National Institute for Materials Science, 1-1 Namiki, Tsukuba 305-0044, Japan}

\author{Takashi Taniguchi}
\affiliation{National Institute for Materials Science, 1-1 Namiki, Tsukuba 305-0044, Japan}

\author{Christian Sch{\"o}nenberger}
\affiliation{Department of Physics, University of Basel, Klingelbergstrasse 82, 4056 Basel, Switzerland}
\affiliation{Swiss Nanoscience Institute, University of Basel, Klingelbergstrasse 82, 4056 Basel, Switzerland}

\date{\today}%

\maketitle

\section*{Materials and Methods}

In the following section we outline the fabrication process of the samples shown in the main text.  We highlight process steps that include heating of the device (\textcolor{red}{\textbf{!!}}), which likely promotes the diffusion of Pd into the WTe$_2$ crystal. 

\begin{enumerate}
\item Palladium bottom contacts.
\begin{itemize}
\item Standard e-beam evaporation of \SI{3}{\nano \meter} titanium (Ti) and \SI{15}{\nano \meter}$^{\ast}$ palladium (Pd) on p-doped Si/SiO$_2$ substrates.
\begin{itemize}
\item[$\ast$] The thickness of Pd with \SI{15}{\nano \meter} was found to provide good superconducting contact while at the same time ensures proper encapsulation of WTe$_2$ by hBN.  A problem arising with increasing Pd thickness is that the covering hBN flake is not sealing off the area around the bottom contacts well enough, such that gaps open along the bottom contacts through which oxygen can enter and oxidize the WTe$_2$ flake.
\end{itemize}
\item Lift-off in \SI{50}{\degree C} acetone.
\end{itemize}

\item Exfoliation of hexagonal boron nitride (hBN).
\begin{itemize}
\item Mechanical exfoliation of hBN on a Si/SiO$_2$ waver with \SI{285}{\nano \meter} oxide using adhesive tape. 
\item Choosing clean flakes of appropriate size with a thickness between \SI{10}{\nano \meter} and \SI{30}{\nano \meter} according to optical contrast \cite{Blake2007}.
\end{itemize}

\item Exfoliation of WTe$_2$ in an inert N$_2$ atmosphere to avoid degradation \cite{Liu2015, Ye2016, Hou2020}. Oxygen level inside the glovebox below $<$ 1 ppm.  
\begin{itemize}
\item Preparation of Si/SiO$_2$ substrates in an oxygen plasma at \SI{30}{\watt} for \SI{5}{min}.  Quick transfer into the glovebox. 
\item Mechanical exfoliation of WTe$_2$ with an adhesive tape and transfer onto a polydimethylsiloxane (PDMS) stamp.
\item PDMS stamp and Si/SiO$_2$ substrate are brought in contact. 
\item Heating of the package for \SI{5}{min} at \SI{120}{\degree C} and full cool-down for \SI{15}{\minute}. 
\item Separation of the PDMS stamp and the substrates. 
\item Selection of elongated WTe$_2$ flakes along the crystallographic a-axis \cite{Zhao2015}. Desired thickness of the flakes $\sim$ \SI{10}{\nano \meter} to \SI{30}{\nano \meter}.
\end{itemize} 

\item Assembly of the vdW hetero-structure via the dry transfer technique \cite{Zomer2014}.
\begin{itemize}
\item Fabrication of the polycarbonate (PC)/PDMS stamp.
\item Pick-up of the selected hBN and afterwards WTe$_2$ flakes at T = \SI{80}{\degree C}.
\item Alignment of the stack with Pd bottom contacts and bringing in stack in contact. 
\item[\textcolor{red}{\textbf{!!}}] Release of the PC/PDMS stamp from the stack by heating the package to \SI{155}{\degree C} for $\sim$ \SI{10}{\minute}.
\item Removal of remaining PC residues in dichloromethane for \SI{1}{\hour}.
\end{itemize}

\item Contacting the stack: Depending on the transport experiment, contact to WTe$_2$ is either made through the Pd bottom contacts or through superconducting contacts made out of aluminium (Al) or molybdenum-rhenium (MoRe). 

\begin{itemize}
\item[\textcolor{red}{\textbf{!!}}] Spin-coating the device with PMMA and successive bake-out at T = \SI{180}{\degree C} for \SI{3}{\minute}.
\item Standard e-beam lithography to pattern the leads and bonding pads.
\item Normal Pd contacts: 
\begin{itemize}
\item e-beam evaporation of $\sim$ \SI{100}{\nano \meter} Au for the leads to contact the Pd bottom contacts.
\item Lift-off in \SI{50}{\degree C} acetone. 
\end{itemize}
\item SC contacts:
\begin{itemize}
\item CHF$_3$/O$_2$ plasma etching$^{\ast \ast}$ in a RIE system to open up the top hBN.
\begin{itemize}
\item[$\ast \ast$] The etching process of the samples was calibrated such that the covering top hBN is safely removed and WTe$_2$ revealed underneath. Calibration of the etch rate for WTe$_2$ is more complex, as bare flakes oxidize and potentially behave differently compared to unoxidized flakes. For this reason, the remaining WTe$_2$ crystal on top of PdTe$_x$ after etching varies slightly between the different samples, as can also be seen from comparison of figs. ~2~a) with ~3~a) in the main text.

Possibly,  the MoRe contact could be separated from PdTe$_x$ by a thin WTe$_2$ layer, forming an additional Josephson junction. However, in both curves,   dV/dI and V(I) in figs.~4~b) and 4~c) of the main text, we observe a zero resistance branch at low bias values. For this reason, the interface between the external contact and the sample does not contribute any parasitic resistance to the junction and a potential additional Josephson junction does not limit the performance of the sample or change the interpretation of the data.
\end{itemize}
\item Transfer of the sample to the sputter machine.  Ignition of an argon (Ar) plasma at \SI{50}{\watt} for \SI{1}{\minute} to clean the contact area.
\item \SI{100}{\nano \meter} thick sputtering of the superconducting leads and bond pads made out of Nb or MoRe.
\end{itemize}
\item Lift-off in \SI{50}{\degree C} acetone. 
\end{itemize}

\end{enumerate}

\section*{Pd Diffusion during the stacking process}

An open question remains when Pd diffuses from the bottom contact into the WTe$_2$ crystal on top.  We have highlighted (\textcolor{red}{\textbf{!!}}) process steps during device fabrication that include heating and could facilitate the diffusion process.  In the following we provide evidence that the diffusion of Pd into WTe$_2$ must happen already during the stacking process of the device. 

\begin{figure}[h!]
    \centering
    \includegraphics[width=0.6\linewidth]{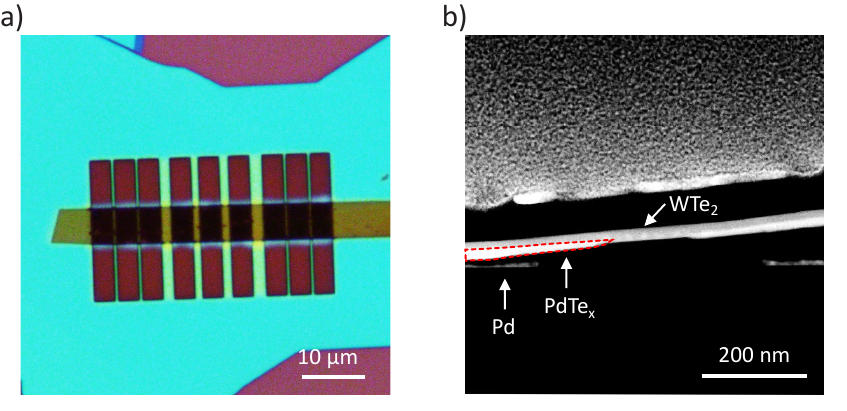}
    \caption{\textbf{Diffusion of Pd during stacking process of the van der Waals (vdW) stack. }  a) Optical image of the finished vdW stack after PC removal. WTe$_2$ is lifted from the Pd bottom contacts, as seen by the color change of hBN in the large spacing between Pd bottom contacts.  b) HAADF image of the same device, showing the WTe$_{2}$ being lifted off the Pd bottom contacts. A swollen PdTe$_x$ crystal structure has visibly formed inside the WTe$_2$ crystal, implying that Pd diffusion must happen already during the PC release at $T =$ \SI{155}{\degree C} after the stacking process. }%
    \label{fig:Diffusion}
\end{figure} 

Fig.~\ref{fig:Diffusion}~a) shows an optical image of the stack after the removal of PC but before etching and deposition of the superconducting contacts. The color change of hBN in between the Pd bottom contacts indicates that the WTe$_2$ flake is lifted off the bottom contacts at this stage.  Fig.~\ref{fig:Diffusion}~b) presents a HAADF STEM image of the same stack,  taken along a JJ. Visible are the two Pd bottom contacts at the left and right end of the image, with the WTe$_2$ spanning across them above.  The WTe$_2$ flake is indeed lifted off the Pd bottom contacts, yet a PdTe$_x$ diffusion layer has clearly formed inside the crystal. For this reason Pd must have diffused already during the stacking process,  when the finished stack is being released on top of the bottom contacts and heated up to $T = $ \SI{155}{\degree C} for $\sim$ \SI{10}{\minute}.  

\section{Horizontal Pd diffusion in the remaining WTe$_2$}

\begin{figure}[h!]
    \centering
    \includegraphics[width=0.7\linewidth]{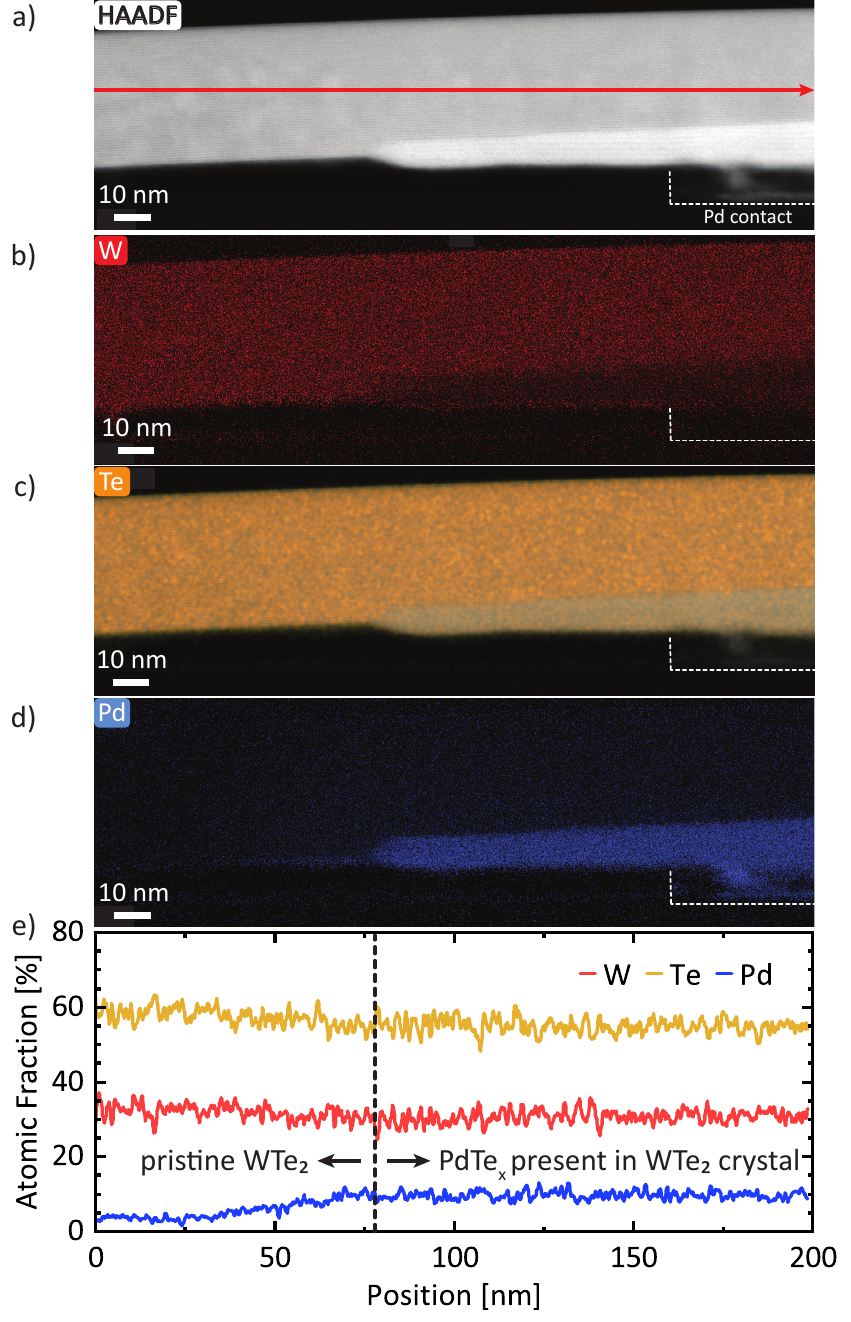}
    \caption{\textbf{Horizontal diffusion of Pd inside the pristine WTe$_2$ crystal.}  a) - d) HAADF image and EDX spectra for the elements W, Te and Pd.  The extend of the former Pd bottom contact is outline by the white dashed line. e) Horizontal line-cut through the EDX spectra taken along the direction of red arrow in a) inside the pristine WTe$_2$ cyrstal above the PdTe$_x$ layer.  The extend of the PdTe$_x$ layer is marked by a vertical dashed line.  The diffused Pd concentration inside pristine WTe$_2$ decreases close to zero at $\sim$ \SI{50}{\nano \meter} beyond the PdTe$_x$ diffusion layer.}%
    \label{fig:EDX horizontal}
\end{figure} 

Figures~\ref{fig:EDX horizontal}~a) through d) present the HAADF image and the EDX spectra for W, Te and Pd of the WTe$_2$ crystal at the edge of the Pd bottom contact, which is indicated by white dashed lines. Figure~\ref{fig:EDX horizontal}~e) plots the atomic fraction in \%  in crystallographically unchanged WTe$_2$, taken along the line cut indicated by the red arrow in fig.~\ref{fig:EDX horizontal}~a). The vertical dashed line in the plot indicates the extension of PdTe$_x$. Beyond this diffusion layer, the concentration of remaining Pd doping in WTe$_2$ decays close to zero within $\sim$ \SI{50}{\nano \meter}.

\section*{STEM and EDX Analysis}

\begin{description}

\item[STEM and EDX analysis of Figs.~1 and 2 of the main text] The TEM-sample preparation has been carried out in a FEI Helios 660 G3 UC FIB/SEM-system. The preparation site was covered with a double layer of platinum in order to protect the layers of interest from any ion induced damage. The first Pt-layer was deposited by means of electron induced deposition at a beam energy of 3 keV and a beam current of 800 pA. This was followed by a second layer of Pt deposited by means of ion induced deposition at a beam energy of 30 kV and a beam current of 230 pA. The specimen cutting and polishing was carried out in the FIB at a beam energy of 30 kV and beam currents ranging from 47 nA down to 80 pA. After reaching a sample thickness of $<$100 nm, the sample was cleaned first at beam energies of 5 kV (beam current 41pA) and finally at 2 kV (beam current 23 pA).
The transfer of the sample from the FIB to the TEM was carried out under argon atmosphere. The imaging of the TEM specimen was carried out in a FEI Titan Themis 3510. The machine was operated in the STEM-mode at a beam energy of 300 keV. The EDX data was post-processed using an average filter of six pixels (pre-filter) in the Velox software. The extracted EDS line scan data was further averaged over 50 neighbouring pixels for each measurement point. 

\item[STEM and EDX analysis of Fig.~3 of the main text] The TEM-sample preparation has been carried out in a FEI Helios NanoLab 650 DualBeam-system. The preparation of the lamella was similar to the already described preparation of Figs. 1 and 2. The imaging of the TEM specimen was carried out in a JEOL JEM-F200. The machine was operated in the STEM-mode at a beam energy of 200 kV. The EDX data were analyzed with the Analysis Station software, thereby the data has not been post-processed.

\end{description}

\section*{Sample Specifications}

Table \ref{Tab1} summarizes the samples used in the main text.  In all samples, WTe$_2$ is placed Pd bottom contacts. 

\begin{table}[h!]
\centering
 \begin{tabular}{|c | c | c | c | c|} 
 \hline
 Sample & Appearance & $d_{WTe_{2}}$ [nm] & $d_{hBN}$ [nm] & SC contact, thickness of material \\ [0.5ex] 
 \hline
 EDX EMPA & Fig.~1, 2 & 35.0 & 12.8 & Nb, \SI{100}{\nano \meter} \\ 
 EDX 210317 & Fig.~3 a) & 23.0 & 53.0 & MoRe, \SI{100}{\nano \meter} \\
 EDX 210805 & Fig.~3 b) & $\sim$\,20* & $\sim$\,25* & No SC, contact through Pd bottom contacts \\
 T 201022 & Fig.~4~b), c) & 21.9 & 24.5 & MoRe, \SI{80}{\nano \meter} \\ 
 T Pd & Fig.~4~d) & $\sim$\,10* & $\sim$\,25* & No SC, contact through Pd bottom contacts\\ [1ex]
 \hline
 \end{tabular}
 \caption{Summary of samples used for the main text. Thickness values of the flakes were determined by AFM. Values highlighted by * were determined through the optical contrast of the flakes \cite{Blake2007}. }
 \label{Tab1}
\end{table}

\newpage

\bibliography{bib_supp}